\documentclass{article}
\usepackage{amssymb,amsfonts,amsmath}

\newcommand{\dsp}{\displaystyle}
\newcommand{\txt}{\textstyle}
\newcommand{\as}{\mathrm{as}}
\newcommand{\al}{\alpha}

\newcommand{\ep}{\epsilon}
\newcommand{\vep}{\varepsilon}
\newcommand{\g}{\gamma}
\newcommand{\ka}{\kappa}
\newcommand{\la}{\lambda}
\newcommand{\n}{\nabla}
\newcommand{\p}{\psi}
\newcommand{\w}{\omega}
\newcommand{\W}{\Omega}
\newcommand{\adv}{\mathrm{adv}}
\newcommand{\out}{\mathrm{out}}
\newcommand{\Hc}{\mathcal{H}}
\newcommand{\K}{\mathcal{K}}
\newcommand{\D}{\partial}
\newcommand{\m}{d\mu}
\newcommand{\dl}{d^2l}
\newcommand{\ov}{\overline}
\newcommand{\V}{\dot{V}}
\newcommand{\cc}{\mathrm{compl.\,conj.}}
\newcommand{\con}{\mathrm{const}}
\DeclareMathOperator{\id}{id} \DeclareMathOperator{\Rp}{Re}
\DeclareMathOperator{\Ip}{Im}
\newcommand\df{:=}

\begin{document}

\title{Asymptotic algebra of quantum electrodynamics}
\author{Andrzej Herdegen%
\thanks{e-mail: herdegen@th.if.uj.edu.pl}\\
{\it Institute of Physics, Jagiellonian University,}\\
{\it Reymonta 4, 30-059 Cracow, Poland}}
\date{}

 \maketitle
\vspace{-.2cm}

\begin{center}
{\sl\textsf{Dedicated to Andrzej Staruszkiewicz on the occasion of
his 65th birthday}}
\end{center}
\vspace{.3cm}
\begin{abstract}
 The Staruszkiewicz quantum model of the long-range structure in
 electrodynamics is reviewed in the form of a Weyl algebra. This
 is followed by a personal view on the asymptotic structure of
 quantum electrodynamics.
\end{abstract}

\section{Introduction}\label{int}

To write on the occasion of the $65^\mathrm{th}$\! birthday of
Professor Andrzej Starusz\-kiewicz is a great honor. When I think
of all the years I have known him I realize that it is at the same
time precisely 30 years ago that I attended, as a first year
student, his lectures in linear algebra and geometry. To many of
us then he was, and remained for all the years of our physics
studies, the most impressive and original teacher. One of the
pictures many of us cherish in our minds is the scene in which he
tries to demonstrate to us that a circle is nothing else than an
interval which has been closed up, using for the purpose, not
quite successfully one must say, the pointing stick he happened to
have in his hand. Anyway, from that moment on I know the
difference between the topology of a line and that of a circle.

This difference, as it happens, becomes prominent in
Staruszkiewicz's quantum theory of the infrared degrees of freedom
of electrodynamics (more on that below). The theory itself is
perhaps the most evident testimony to what some of us had the
opportunity to discover later on: that Sta\-rusz\-kie\-wicz's
appeal as a teacher reflected the inherent originality of his
thinking on physics, and beyond. The author of these words counts
among those whose style of physics-making was greatly influenced,
albeit sometimes in polemics, by Professor Staruszkiewicz. More
than that, Staruszkiewicz's ideas on the long-range properties of
quantum electrodynamics were among those, which aroused my own
steady interest in the field. This encourages me to use this
opportunity to sketch a pedagogically oriented review of the
Staruszkiewicz's model, as I see it, and to follow this by my own
view of the asymptotic algebraic structure of quantum
electrodynamics. It should be acknowledged, I hope that Professor
Staruszkiewicz will agree, that the issues of infrared structure
of electrodynamics remain to be controversial. Thus it is only
natural that in addition to important common points in our views
there will be other on which we differ.

The problem we want to address is the following:

What are the consequences of the long-range character of the
electromagnetic interaction for the algebraic structure of the
quantum theory of radiation and charged particles? What is the
algebraic formulation of Gauss' law, and can it implement the
charge quantization?

The approaches to this question summarized here start from
concrete structures rather than from general assumptions on the
desirable properties of the electromagnetic theory. To place this
work in a wider context we start with some general remarks.

Quantum electrodynamics shares many properties (and difficulties)
with other quantum field-theoretical models. However, its most
interesting ingredients are those, that in our opinion (apparently
shared by Staruszkiewicz) are specific to this theory -- its long
range structure being especially prominent. By this term we mean
the group of properties connected with the masslessness of the
photon, existence and quantization of electric charge and Gauss'
law (see e.g.\ \cite{jr},\,\cite{ms}). The lack of complete,
mathematically sound formulation of quantum electrodynamics is, of
course, an obstacle to conclusive understanding of the long range
structure. This structure, however, needs only low energies and
asymptotic spacetime regions to manifest itself. Therefore, it
only very weakly involves the dynamics of the system, which lends
some support to the hope that understanding this structure does
not presuppose the complete understanding in detail of the
dynamics. This belief lies at the base of the investigation of the
long range structure from the ``axiomatical'' point of view. The
main result of this study may be briefly formulated as follows:
the flux of the electromagnetic field at spacelike infinity is
superselected in irreducible representations of local observables
\cite{bu}. More precisely, if the leading term of the
electromagnetic field is well defined and decays as the Coulomb
field in spacelike directions, then its distribution in spacelike
infinity is fixed in such representations. One can say, therefore,
that this long-range electromagnetic field has a classical
spectrum, and elements of this spectrum (functions of the angles)
label different sectors. Representations of local observables from
different sectors are unitarily inequivalent. In particular,
states differing in total charge value are inequivalent in
consequence of Gauss' law. There are also two other important
consequences of this superselection structure: (i) in charged
sectors the Lorentz group is spontaneously broken
\cite{fms},\cite{buch}; and (ii) the contribution of a charged
particle to the spectrum of squared four-momentum (the mass
spectrum squared) is not point-like. The former means, that
although the Lorentz transformations of the observables are
defined, they cannot be obtained by the action of a unitary
representation of the Lorentz group in the representation space of
the charged state. The consequence of the latter is that a charged
particle, being accompanied by its electromagnetic field, is an
object far more complex then a ``bare'', neutral particle. This
fact is referred to as ``the infraparticle problem''
\cite{sch},\cite{buch}. The concept of an elementary particle has
to be revised, in consequence, to be applicable to an electron.
Several suggestions for such revisions has been formulated, among
them the proposal by Buchholz \cite{bw} to use weights on the
C$^*$-algebra of observables for the generalization of the
particle concept seems to be the most far-reaching.

The great value of the axiomatic approach to the quantum field
theory problems lies in discovering strict logical connections
between the expected fundamental features of the underlying
structure on the one hand, and the interpretational (physical)
properties of a theory (model) based on it. Among the basic
postulates is the locality of observables: that each observable
quantity may be measured locally in a compact subset of spacetime,
or be derived as a limit of such local quantities (see
\cite{haa}). However, physics deals with idealizations, and one
could ponder whether in the case of electrodynamics, which
includes constraints with nonlocal consequences (the Gauss law),
we would not learn something by enlarging the scope of the
admitted observables by some ``variables at infinity''.

In the two models summarized here such variables appear in a
natural way. Also, both models include one variable of the phase
type (circle topology), whose presence leads to the charge
quantization. In other respects they differ. Staruszkiewicz
considers the spacelike limit of classical electromagnetic fields
and quantizes the resulting structure. The model has the advantage
of (relative) simplicity, and in fact is probably a minimal field
theoretical structure containing the Coulomb field among its
variables. This is sufficient for Staruszkiewicz's main objective,
which is to look for the justification of the actual value of
elementary charge (or rather, one should say, the dimensionless
fine structure constant). Formulation of the model is given in
\cite{sta1}, additional discussion of the motivation may be found
in \cite{sta2}.

My own aim is different, and the intention is to stay closer to
the standard analysis. The object sought is the algebra of the
asymptotic fields, in the causal, ``in'' or ``out'' sense. If we
had a complete quantum theory at our disposal, we could try to
obtain the algebra in the respective limits. Lacking this one
tries to make a guess based on intuitions formed by simpler
quantum models. Perturbational quantum electrodynamics treats the
asymptotic fields as uncoupled. This, however, is a wrong
idealization, not respecting Gauss' law. Our method is to quantize
the causal limits of the classical fields: timelike for matter and
lightlike for electromagnetic fields. For separate free fields
this reproduces the usual quantization (for the electromagnetic
case: as considered at null infinity by Bramson and Ashtekar
\cite{bras}). For interacting fields, however, a remnant of
interaction survives, which correctly incorporates the
consequences of Gauss law, and which is truly nonlocal. This may
be interpreted as some form of ``dressing'' of a charged particle,
and thus has relations with earlier works by Kulish and Fadeev
\cite{kf}, Fr\"ohlich \cite{fr}, Zwanziger \cite{zw}, and others.
However, here we are able to obtain a closed algebra which may be
expected to have fairly universal features adequately
incorporating the long range structure. A~formulation and
discussion of the model is to be found in \cite{her1}.

We use physical units in which $\hslash=1$, $c=1$.

\section{Asymptotic fields at spacelike infinity}\label{space}

We start with a discussion of the spacelike limit of classical
fields. Suppose that $A(x)$ is a classical field satisfying the
wave equation. Its Fourier representation is then given by
\begin{equation}\label{free}
 A(x)=\frac{1}{\pi}\int a(k)\delta(k^2)\vep(k^0)e^{-ix\cdot
 k}\,d^4k\,,
\end{equation}
where $\delta$ is the Dirac delta function and $\vep$ is the sign
function. If $A_b(x)$ is an electromagnetic vector potential in
Lorentz gauge of a free electromagnetic field, then $a_b(k)$ is a
vector function satisfying $k\cdot a(k)=0$ on the light-cone, and
the reality of $A_b(x)$ is equivalent to
\begin{equation}\label{reality}
 \ov{a_b(k)}=-a_b(-k)\,.
\end{equation}
If $a_b(k)$ is a smooth function then $A_b(x)$ decreases rapidly
in spacelike directions. However, as is well-known, the spacelike
decay of the actual radiation fields produced in real processes is
determined by the rate of decrease of the Coulomb fields of the
sources. Thus one considers a wider class of potentials, those
with well-defined spacelike scaling limit:
\begin{equation}\label{sppot}
 A^\mathrm{as}_b(x)\df\lim_{\la\to\infty}\la A_b(\la x)\,,\quad
 x^2<0\,,
\end{equation}
which is expressed in terms of the Fourier transform as the
existence of the limit
\begin{equation}\label{aas}
 a^\as_b(k)=\lim_{\mu\searrow0}\mu\, a_b(\mu k)\,.
\end{equation}
Note that both $A^\as_b(x)$ and $a^\as_b(k)$ are homogeneous
functions of degree $-1$.

Before proceeding further let us remind the reader that if $f(l)$
is a function of a future-pointing null vector $l$, homogeneous of
degree $-2$, written as \mbox{$f(l)=f(l^0,\vec{l})$} in a given
Minkowski basis, then the following integral
\begin{equation}
 \int f(l)\,\dl\df\int f(1,\hat{l})\,d\W(\hat{l})
\end{equation}
is Lorentz invariant, i.e.\ independent of the basis (here
$\hat{l}$ is a unit vector in $3$-space and $d\W(\hat{l})$ is the
solid angle measure). We also note for later use that the
differentiations of functions on the cone in tangent directions
may be conveniently expressed by the application of the operators
\begin{equation}
 L_{ab}\df l_a\D_b-l_b\D_a\,,\qquad\text{where}
 \quad \D_a\df\frac{\D}{\D l^a}\,,
\end{equation}
 and that
\begin{equation}\label{lint}
 \int L_{ab}f(l)\,\dl=0\,.
\end{equation}
Note that also the operator $l\cdot\D$ is intrinsically defined on
the lightcone, as $l_a\,l\cdot\D=l^cL_{ac}$. Furthermore, if
$h(l)$ is a regular function on the cone (except, possibly, its
tip) and for the sake of differentiation one extends it in a
regular, but otherwise arbitrary way to a neighborhood of the cone
(outside the tip), then one shows that on the cone itself one has
\begin{equation}
 [L_{ab}L_c{}^b+L_{ac}]h=
 [l_al_c\D^2-(l_a\D_c+l_c\D_a+g_{ac})l\cdot\D]h\,.
\end{equation}
As the operator on the l.h.\ side is intrinsically defined on the
cone, the same must be true for the r.h.\ side. In particular, if
$h(l)$ is homogeneous of degree~$0$, then on the cone one has
\begin{equation}
 [{}^*\!L_{cb}{}^*\!L_a{}^b+L_{ac}]h
 =[L_{ab}L_c{}^b+L_{ac}]h=l_al_c\D^2h\,,
\end{equation}
where star denotes the dual of an antisymmetric tensor. This shows
that in this case the expression $\D^2h(l)$ determines a
homogeneous function of degree $-2$ intrinsically on the cone,
which in each Minkowski basis may be represented by
\begin{equation}
 \D^2h=(l^0)^{-2}{}^*\!L_{0b}{}^*\!L_0{}^bh\,.
\end{equation}
In a similar way one shows that for two functions $h_1(l)$ and
$h_2(l)$ homogeneous of degree $0$ one has
\begin{equation}
 {}^*\!L_{cb}h_1{}^*\!L_a{}^bh_2=L_{ab}h_1L_c{}^bh_2
 =l_al_c\,\D h_1\cdot\D h_2\,.
\end{equation}
Thus $\D h_1\cdot\D h_2$ is intrinsically defined on the cone, and
in each Minkowski basis there is
\begin{equation}
 \D h_1\cdot\D h_2=(l^0)^{-2}{}^*\!L_{0b}h_1{}^*\!L_0{}^bh_2\,.
\end{equation}
Taking into account that ${}^*\!L_{0b}l^0=0$ and integrating by
parts with the use of (\ref{lint}) one has now
\begin{equation}
 \int\D^2h_1\,h_2\,\dl=
 -\int\D h_1\cdot\D h_2\,\dl=\int h_1\D^2 h_2\,\dl\,.
\end{equation}

We can now return to the discussion of the asymptotic field.
Calculating the asymptotic spacelike limit for the Fourier
representation one shows that it becomes in this limit
\begin{equation}\label{aspot}
 A^\as_b(x)=\frac{-i}{2\pi}\int\frac{a^\as_b(l)}{x\cdot l-i0}\,\dl
 +\cc\,,
\end{equation}
which yields the asymptotic electromagnetic field
\begin{equation}\label{asfield}
 F^\as_{bc}(x)=\frac{i}{2\pi}\int
 \frac{l_ba^\as_c(l)-l_ca^\as_b(l)}{(x\cdot l-i0)^2}\,\dl
 +\cc
\end{equation}
(both $(x\cdot l-i0)^{-1}$ and $(x\cdot l-i0)^{-2}$ are
well-defined homogeneous distributions). We stress that here, and
throughout the paper, $l$ always denotes a \emph{future-pointing}
null vector. Now, one can show that there exist unique up to
additive constants, homogeneous of degree $0$ complex functions
$a(l)$ and $b(l)$ such that
\begin{equation}
 l_ba^\as_c(l)-l_ca^\as_b(l)=L_{bc}a(l)-{}^*\!L_{bc}b(l)
\end{equation}
-- this follows from homogeneity of degree $-1$ of $a^\as_b(l)$
and its orthogonality to~$l^b$, and can be shown most easily with
the use of spinor formalism. We can thus separate $F^\as_{ab}$
into two parts:
\begin{equation}
 F^\as_{ab}=F^\mathrm{E}_{ab}+F^\mathrm{M}_{ab}\,,
\end{equation}
where
\begin{gather}
 F^\mathrm{E}_{ab}(x)=\frac{i}{2\pi}\int
 \frac{L_{ab}a(l)}{(x\cdot l-i0)^2}\,\dl
 +\cc\,,\label{asfE}\\
 {}^*\!F^\mathrm{M}_{ab}(x)=\frac{i}{2\pi}\int
 \frac{L_{ab}b(l)}{(x\cdot l-i0)^2}\,\dl
 +\cc\label{asfM}
\end{gather}
Using this form one finds that
$F^\mathrm{E}_{[ab}x_{c]}={}^*\!F^\mathrm{M}_{[ab}x_{c]}=0$. This
follows from the identity
 $x_c(x\cdot l-i0)^{-2}=-\D_c(x\cdot l-i0)^{-1}$
 and the following transformations of the integral
\begin{equation*}
 \int L_{[ab}\,a\,\D_{c]}\frac{1}{x\cdot l-i0}\,\dl
 =\int\D_{[b}a\, L_{ac]}\frac{1}{x\cdot
 l-i0}\,\dl
 =-\int \frac{L_{[ac}\D_{b]}a}{x\cdot
 l-i0}\,\dl=0\,,
\end{equation*}
and similarly for $b(l)$. In consequence
\begin{equation}\label{fk}
 F^\mathrm{E}_{ab}(x)=x_aK^\mathrm{E}_b(x)-x_bK^\mathrm{E}_a(x)\,,\quad
 {}^*\!F^\mathrm{M}_{ab}(x)=x_aK^\mathrm{M}_b(x)-x_bK^\mathrm{M}_a(x)\,,
\end{equation}
where
\begin{equation}\label{k}
 K^\mathrm{E}_a(x)=\frac{1}{x^2}\,x^cF^\mathrm{E}_{ca}\,,\quad
 K^\mathrm{M}_a(x)=\frac{1}{x^2}\,x^c{}^*\!F^\mathrm{M}_{ca}\,.
\end{equation}
This form shows that $F^\mathrm{E}_{ab}$ and $F^\mathrm{M}_{ab}$
are fields of electric and magnetic type respectively: one can
check directly that the long range tail produced by scattered
electric charges is of type $F^\mathrm{E}_{ab}$; by duality,
$F^\mathrm{M}_{ab}$ would appear in scattering of magnetic
monopoles.

Thus, being interested in the actual electrodynamics, we do not
need to include long-range fields of the magnetic type in the
theory, and from now on we assume that
\begin{equation}
 F^\mathrm{M}_{ab}=0\,,\quad \text{that is}\quad b= 0\,.
\end{equation}
In that case we have
\begin{equation}\label{al}
 l_ba^\as_c(l)-l_ca^\as_b(l)=L_{bc}a(l)\,,
\end{equation}
so
\begin{equation}
 a_b^\as(l)=\D_ba(l)+l_b\al(l)\,,
\end{equation}
where $a(l)$ has been extended for the sake of differentiation to
a homogeneous function in a neighborhood of the lightcone, and
$\al(l)$ is a homogeneous function of degree $-2$. The second term
does not contribute to the field $F^\as_{ab}$, so it must yield a
gauge term in the potential, and indeed:
\begin{equation}\label{homgauge}
 \frac{-i}{2\pi}\int\frac{l_b\al(l)}{x\cdot l-i0}\,\dl
 =\n_b\frac{-i}{2\pi}\int\al(l)
 \log\Big[\frac{x\cdot l-i0}{t\cdot l}\Big]\,\dl\,,
\end{equation}
where $t$ is any future-pointing unit timelike vector and
$\n_b\df\D/\D x^b$. However, we note that the omission of this
term does not leave an unambiguously defined gauge invariant
expression for the asymptotic potential. Although $a$ and $\D^2a$
are intrinsically defined on the cone, the expression $\D_ba$ is
not, and depends on the choice of homogeneous extension of $a$ to
the neighborhood of the cone: two different homogeneous extensions
yield two $\D_ba(l)$'s differing by a term of the form
$l_b\beta(l)$.\footnote{For instance, for the homogeneous function
$f(l)=l^2/(t\cdot l)^2$ we have on the cone: $f=0$, but
$\D_af(l)=2l_a/(t\cdot l)^2$.} This corresponds to a change of
gauge in $A^\as_b$, therefore not all information on the potential
$A^\as_b$ may be encoded in the light-cone function $a(l)$.

The electromagnetic field $F^\as_{ab}$ is most compactly expressed
with the use of Eqs.\ (\ref{fk}) and (\ref{k}). We can now
identify $F^\as_{ab}=F^\mathrm{E}_{ab}$ and write
$K_a=K^\mathrm{E}_a$, so by the homogeneity properties we have
\begin{equation}\label{ks}
 x^2 K_a(x)=x^cF^\as_{ca}(x)=\n_a[-x\cdot A^\as(x)]
 =\frac{1}{e}\n_aS(x)\,,
\end{equation}
where $e$ is the elementary charge, and following Staruszkiewicz
we have denoted
\begin{equation}
 S(x)=-e\,x\cdot A^\as(x)\,.
\end{equation}
For any future-pointing unit timelike vector $t$ there is
\begin{equation}
 \frac{x\cdot a^\as(l)}{x\cdot l-i0}
 =\D a(l)\cdot\D\log\Big[\frac{x\cdot l-i0}{t\cdot l}\Big]
 +\frac{t\cdot a^\as(l)}{t\cdot l}\,,
\end{equation}
so using (\ref{aspot}) one finds that
\begin{gather}
 S(x)=\frac{e}{2}\int\D^2\Rp a(l)\,\vep(x\cdot l)\,\dl
 +\frac{e}{\pi}\int\D^2\Ip a(l)
 \log\Big[\frac{|x\cdot l|}{t\cdot l}\Big]\,\dl+S_t\,,\label{phase}\\
 S_t\equiv-\frac{e}{\pi}\int
 \frac{t\cdot \Ip a^\as(l)}{t\cdot l}\,\dl\,.\label{phaset}
\end{gather}
This scalar function, homogeneous of degree zero,  contains the
whole information on the field $F^\as_{ab}(x)$, and in addition
has an additive constant $S_t$ not contributing to this field.
This constant is both gauge- and $t$-dependent:
\begin{equation}\label{stgauge}
 \text{if}\quad \tilde{a}^\as_b(l)=a^\as_b(l)+l_b\beta(l)\,,\quad
 \text{then}\quad \tilde{S}_t=S_t-\frac{e}{\pi}\int\beta(l)\,\dl\,,
\end{equation}
and if $t'$ is another future-pointing unit timelike vector, then
\begin{equation}
 S_{t'}=S_t+\frac{e}{\pi}\int\D^2\Ip a(l)
 \log\Big[\frac{t'\cdot l}{t\cdot l}\Big]\,\dl\,.
\end{equation}
The last transformation property confirms that the $t$-dependence
of the formula (\ref{phase}) is spurious. On the other hand, the
whole function $S(x)$ also undergoes the gauge
transformation:\footnote{We should acknowledge here that
Staruszkiewicz regards $S(x)$ as gauge-independent. This is a
consequence of his apparent treating $\D_ba(l)$ as an
unambiguously defined quantity. Note also, that in general the
contraction of the gauge term (\ref{homgauge}) with $x^b$ does not
vanish.}
\begin{equation}
 \tilde{S}(x)=S(x)-\frac{e}{\pi}\int\beta(l)\,\dl\,.
\end{equation}

We stated above that not the whole information on $a^\as_b(l)$ is
contained in~$a(l)$. However, as it turns out, the freedom of
adding a constant to $a(l)$ may be used to choose this function so
as to contain the whole information on $S(x)$. Namely, given
$a^\as_b(l)$, a special solution of Eq.\,(\ref{al}) for $a(l)$ may
be shown to be
\begin{equation}
 a(l)=\frac{1}{4\pi}\int\frac{l\cdot a^\as(l')}{l\cdot
 l'}\,\dl'\,.
\end{equation}
This solution has the following remarkable property: for each unit
timelike vector $t$ there is
\begin{equation}
 \int\frac{a(l)}{(t\cdot l)^2}\,\dl
 =\int\frac{t\cdot a^\as(l)}{t\cdot l}\,\dl\,,
\end{equation}
so with this choice, which will always be assumed from now on, we
have
\begin{equation}
 S_t=-\frac{e}{\pi}\int\frac{\Ip a(l)}{(t\cdot l)^2}\,\dl\,.
\end{equation}
The function $S(x)$ is now seen to be determined completely and
uniquely by $\D^2\Rp a(l)$ and $\Ip a(l)$.

\section{Staruszkiewicz's model}\label{stam}

At this point one observes that $S(x)$ satisfies the wave equation
\begin{equation}\label{wave}
 \Box S(x)=0\,,
\end{equation}
and that Eq.\,(\ref{phase}) almost gives the most general function
homogeneous of degree zero satisfying this equation.\footnote{Eq.\
(\ref{wave}) together with the homogeneity are equivalent to the
wave equation on the hyperboloid $x^2=-1$.\label{f}} The
reservation ``almost'' is due to the fact that in place of
$\D^2\Rp a(l)$ one can have an arbitrary function $c(l)$
homogeneous of degree $-2$. This makes a difference of only one
degree of freedom. Namely, if $t$ is any timelike, unit,
future-pointing vector, and one denotes
\begin{equation}
 c_t(l)=c(l)-\frac{\int c(l')\,\dl'}
 {4\pi(t\cdot l)^2}\,,\quad
 \text{then}\quad \int c_t(l)\,\dl=0\,.
\end{equation}
But each function satisfying the last equation may be represented
as a result of applying $\D^2$ to a homogeneous function of degree
$0$, so the only quantity lacking from (\ref{phase}) is $\int
c(l)\dl$. Following Staruszkiewicz we now add this degree of
freedom. Thus we:
\begin{equation}\label{repl}
 \text{replace}\quad\D^2\Rp a(l)\to -\frac{1}{2\pi}c(l)\,, \quad
 \text{and denote}\quad \Ip a(l)\equiv-\frac{1}{4} D(l)\,,
\end{equation}
where the choice of constants is a mere convention. Our function
$S(x)$ becomes now
\begin{gather}
 S(x)=-\frac{e}{4\pi}\int c(l)\,\vep(x\cdot l)\,\dl
 -\frac{e}{4\pi}\int\D^2 D(l)
 \log\Big[\frac{|x\cdot l|}{t\cdot l}\Big]\,\dl+S_t\,,\label{phcom}\\
 S_t=\frac{e}{4\pi}\int\frac{D(l)}{(t\cdot l)^2}\,\dl\,.
 \label{phcomt}
\end{gather}
If thus extended function $S(x)$ is now used in (\ref{ks}) to
determine the asymptotic field $F^\as_{ab}(x)$, then the new
degree of freedom added to $S(x)$ produces a charged field, with
charge given by
\begin{equation}\label{charge}
 Q=\frac{1}{4\pi}\int c(l)\,\dl
\end{equation}
-- this is shown by integrating the flux of electric field over a
sphere.

Staruszkiewicz's model now rests upon two main suppositions: that
one can base a model of the long-range structure on the field
$S(x)$ alone, and that $S_t$ should be interpreted as a phase
variable. For the motivation we refer the reader to the original
papers by Staruszkiewicz. Consider the first supposition. One
looks for a quantization condition for $\hat{S}(x)$ of the form
$[\hat{S}(x),\hat{S}(y)]\propto\id$, where ``hats'' indicate the
quantum versions of these variables. This should be expressible as
$[\hat{c}(l),\hat{D}(l')]\propto\id$. Let $D(l)$ and $c(l)$ be now
classical test functions, homogeneous of degree $0$ and $-2$
respectively, and denote
\begin{equation}\label{cD}
 \hat{c}(D)=\frac{1}{4\pi}\int\hat{c}(l)D(l)\,\dl\,,\quad
 \hat{D}(c)=\frac{1}{4\pi}\int\hat{D}(l)c(l)\,\dl\,.
\end{equation}
Then the only Lorentz-covariant quantization condition, up to a
multiplicative constant on the r.h.\ side, is
\begin{equation}\label{comrel}
 [\hat{c}(D),\hat{D}(c)]=\frac{i}{4\pi}\int D(l)c(l)\,\dl\,\id
\end{equation}
-- the choice of the particular constant will be justified in a
moment. A straightforward calculation with the use of
(\ref{phcom}) yields now
\begin{equation}
 \begin{split}
 [\hat{S}(x),\hat{S}(y)]
 =&i2e^2\vep\Big(\frac{x^0}{\sqrt{-x^2}}
 -\frac{y^0}{\sqrt{-y^2}}\Big)\times\\
 &\times\theta\Big(
 \Big[\frac{x}{\sqrt{-x^2}}-\frac{y}{\sqrt{-y^2}}\Big]^2\Big)
 \frac{x\cdot y}{\sqrt{(x\cdot y)^2-x^2y^2}}\,\id\,,
 \end{split}
\end{equation}
which is the relation obtained by another method by
Staruszkiewicz.\footnote{There is a misprint of a sign on the
r.h.\ side of this relation in \cite{sta1}.} This commutation
relation guarantees causality when restricted to the unit
hyperboloid $x^2=-1$.

Consider now the second supposition, that $\hat{S}_t$ is a phase
variable. Using (\ref{phcomt}), (\ref{charge}) and (\ref{cD}) one
finds that
\begin{equation}
 \hat{S_t}=\hat{D}\Big(\frac{e}{(t\cdot l)^2}\Big)\,,\quad
 \hat{Q}=\hat{c}(1)\,,
\end{equation}
so by (\ref{comrel}) one has
\begin{equation}\label{QS}
 [\hat{Q},\hat{S}_t]=ie\,\id\,.
\end{equation}
The supposition means that in this relation $\hat{S}_t$ should be
used, in fact, in the form $\exp[-i\hat{S}_t]$, and the above
commutation relation should be understood as
\begin{equation}
 \hat{Q}e^{\txt-i\hat{S}_t}=e^{\txt-i\hat{S}_t}(\hat{Q}+e\id)\,.
\end{equation}

The precise formulation of the commutation relations thus obtained
has the following Weyl form derived by the heuristic substitution
\begin{equation}\label{heur}
 W(D)=\exp\big[i\hat{c}(D)\big]\,,\quad
 R(c)=\exp\big[-i\hat{D}(c)\big]\,,
\end{equation}
and by admitting in $R(c)$ only those test functions $c$ for which
there is
\begin{equation}
 n_c\df\frac{1}{4\pi e}\int c(l)\,\dl\in\mathbb{Z}\,.
\end{equation}
The algebraic relations are
\begin{equation}\label{weylst}
 \begin{split}
 &\hspace{-15pt}W(D)W(D')=W(D+D')\,,\quad
 R(c)R(c')=R(c+c')\,,\\
 &W(D)R(c)=\exp\Big\{\dfrac{i}{4\pi}\int
 D(l)c(l)\,\dl\Big\}R(c)W(D)\,,\\
 &\hspace{-25pt}W^*(D)=W(-D)\,,\quad R^*(c)=R(-c)\,,\quad
 W(0)=R(0)=\id\,,
 \end{split}
\end{equation}
which defines an abstract Weyl algebra. To consider a physical
realization of the system one needs a *-representation of this
algebra by operators in a Hilbert space.\footnote{Not to burden
notation we shall keep the same symbol for the operator, as for
the abstract element itself.} Before choosing a particular
representation we make some comments on the structure of the
algebra.

First of all, one should observe that the algebra could be
formulated in terms more directly connected with the spacetime
relations. Namely, for any two homogeneous solutions of the wave
equation (\ref{wave}) the formula
\begin{equation}\label{symplS}
 \{S_1,S_2\}\df\sqrt{(x^0)^2+1}\int
 \big[S_1\n_0S_2-S_2\n_0S_1\big]
 \big(x^0,\sqrt{(x^0)^2+1}\,\hat{x}\big)\,d\W(\hat{x})\,,
\end{equation}
where $\hat{x}$ is a vector on a unit sphere in 3-space and
$d\W(\hat{x})$ is the solid angle measure, defines a symplectic
form conserved under the evolution and independent of the
reference system.\footnote{This is the symplectic form for
solutions of the wave equation on the hyperboloid $x^2=-1$, cf.\
footnote \ref{f}.} On the other hand one can show that
\begin{equation}
 \{S_1,S_2\}=e^2\int\big[c_1D_2-c_2D_1\big](l)\,\dl\,,
\end{equation}
if $S_i(x)$ are represented as in (\ref{phcom}). Thus the initial
values $S(0,\hat{x})$, $\n_0S(0,\hat{x})$ could be used instead of
$c(l)$, $D(l)$ as test fields of the algebra elements. This leads
to relativistic locality of the commutation relations on the
hyperboloid $x^2=-1$, but we do not go in any further details.

Next, we note two important symmetries of the algebra. For
$\lambda\in\mathbb{R}$ we have a group of automorphisms of the
algebra defined by
\begin{equation}
 \g_\la(A)=W(\la)AW(-\la)\,,\quad
 \g_\la\g_{\la'}=\g_{\la+\la'}\,.
\end{equation}
By the basic commutation relations we have
\begin{equation}
 \g_\la\big(W(D)\big)=W(D)\,,\quad
 \g_\la\big(R(c)\big)=e^{i\la n_ce}R(c)\,.
\end{equation}
In representations in which $W(\la)$ is regular we have
$W(\la)=\exp[i\la\hat{Q}]$, where $\hat{Q}$ has the interpretation
of the charge operator. Therefore the automorphism~$\g_\la$ should
be regarded as a (global) gauge transformation. Accordingly, the
algebra of observables is the subalgebra of (\ref{weylst})
consisting of elements invariant under~$\g_\la$, which is
generated by the elements of the form $W(D)R(\D^2F)$ with $F(l)$
homogeneous of degree $0$ (recall that if $n_c=0$ then there
exists such $F$ that $c=\D^2F$). Elements $R(c)$ with $n_c\neq0$
are field variables interpolating between superselection sectors
and creating the charge $n_ce$. This confirms our earlier
statement that $S_t$ is a gauge dependent quantity, which should
not be regarded as an observable. Note, however, that
$R(c)^*R(c')=R(c'-c)$ \emph{is} an observable if $n_c=n_{c'}$, so
sectors are labelled only by charge value. Note, moreover, that
$\g_{(2\pi/e)}=\id$. Thus if the representation of (\ref{weylst})
is irreducible then $\exp[i2\pi\hat{Q}/e]\propto\id$. If in
addition $0$~is in the spectrum of $\hat{Q}$, then the spectrum is
equal to $e\mathbb{Z}$. This leads to the quantization of charge
and justifies the choice of the multiplicative constant in the
quantization condition (\ref{comrel}).

Another symmetry group of the algebra (\ref{weylst}) is the
Lorentz group, which acts on the algebra by the automorphisms
($\Lambda$ is a Lorentz transformation):
\begin{equation}
 \begin{split}
 \al_\Lambda\big[W(D)\big]=W(T_\Lambda D)\,&,\quad
 \al_\Lambda\big[R(c)\big]=R(T_\Lambda c)\,,\\
 \text{where}\qquad
 [T_\Lambda D](l)=D(\Lambda^{-1}l)\,&,\quad
 [T_\Lambda c](l)=c(\Lambda^{-1}l)\,.
 \end{split}
\end{equation}
There is no nontrivial translation symmetry in the algebra.

One looks for representations which have a cyclic vector $\W$
(that is the closure of the linear span of all vectors
$W(D)R(c)\W$ is the whole representation space), in which the
Lorentz symmetry is implementable, i.e.\ there exists a unitary
representation of the Lorentz group $U(\Lambda)$ such that for
each operator $A$ in the representation of the algebra there is
\begin{equation}\label{lor}
 \al_\Lambda(A)=U(\Lambda)AU^*(\Lambda)\,,
\end{equation}
and in which $\W$ is Lorentz-invariant:
\begin{equation}\label{invac}
 U(\Lambda)\W=\W\,.
\end{equation}
A class of such representations may be obtained by the Fock method
(we are not aware of a proof that this exhausts the set of
covariant representations). Assume that the operators of the
observable elements $W(D)$ and $R(\D^2F)$ are regular, that is
there exist selfadjoint $\hat{c}(D)$ and $\hat{D}(\D^2F)$ such
that for $\lambda\in\mathbb{R}$ there is $W(\lambda
D)=\exp[i\lambda\hat{c}(D)]$ and
$R(\lambda\D^2F)=\exp[-i\lambda\hat{D}(\D^2F)]$. Let $\ka$ be any
real positive number. Suppose that in the representation space
there exists a vector $\W_\ka$ which is cyclic and for each $F(l)$
homogeneous of degree $0$ satisfies
\begin{equation}\label{covrep}
 \Big[\sqrt{\ka}\,\hat{c}(F)+\frac{i}{\sqrt{\ka}}\hat{D}(\D^2F)\Big]
 \W_\ka=0\,.
\end{equation}
One shows that these conditions determine a unique (up to a
unitary equivalence) representation. We sketch the proof. Suppose
first that such $\W_\ka$ exists. From the condition (\ref{covrep})
for $F=1$ we have in particular \mbox{$\hat{Q}\W_\ka=0$}.
Moreover, from the commutation relations we get
\begin{equation}
 \hat{Q}\,W(D)R(c)\W_\ka=n_ce\,W(D)R(c)\W_\ka\,.
\end{equation}
Therefore the representation space is
\begin{equation}
 \Hc=\bigoplus_{n\in\mathbb{Z}}\Hc_n\,,\quad\text{where}\quad
 \hat{Q}\Hc_n=ne\Hc_n\,,
\end{equation}
and $\Hc_n$ is the closure of the linear span of vectors
$W(D)R(c)\W_\ka$ with \mbox{$n_c=n$}. It is now easy to see that
all matrix elements of operators $W(D)R(c)$ between arbitrary
vectors from the set $W(D')R(c')\W_\ka$ are reduced with the use
of commutation relations either to zero, or to a matrix element in
the space $\Hc_0$. It is thus sufficient to show the existence and
uniqueness of representation of observable elements $W(D)R(\D^2F)$
in $\Hc_0$. For that purpose for each complex function~$F(l)$
homogeneous of degree $0$ let us denote by $[F]$ its equivalence
class with respect to the addition of a constant. Let $\K$ be the
Hilbert space of such classes with the scalar product
\begin{equation}
 ([F],[G])_\K=\frac{1}{4\pi}\int(-\D F\cdot\D G)\,\dl\,,
\end{equation}
and let $\Hc_0$ be the Fock space based on the ``one-excitation''
space $\K$. Denote by $\W$ the ``Fock vacuum'' vector and by
$d([F])$ the annihilation operator in that Fock space:
\begin{equation}
 d([F])\W=0\,,\qquad \big[d([F]),d^*([G])\big]=([F],[G])_\K\id\,.
\end{equation}
We set $\W_\ka=\W$ and for real $F$
\begin{equation}
 \begin{split}
 \hat{c}(F)|_{\Hc_0}
 &=\frac{1}{\sqrt{2\ka}}\Big\{d([F])+d^*([F])\Big\}\,,\\
 \hat{D}(\D^2F)|_{\Hc_0}
 &=-i\sqrt{\frac{\ka}{2}}\Big\{d([F])-d^*([F])\Big\}\,.
 \end{split}
\end{equation}
It is easy to show that this ensures the correct commutation
relations and that Eq.\,(\ref{covrep}) is now satisfied, so the
existence of the representation is proved. Furthermore, it follows
from (\ref{covrep}) alone that
\begin{equation}\label{vacexp}
 (\W_\ka,W(D)R(\D^2F)\W_\ka)=
 \exp\frac{1}{4}\Big[-\ka^{-1}\|[D]\|^2_\K-\ka\|[F]\|^2_\K
 +i2([D],[F])_\K\Big]\,.
\end{equation}
As by the GNS construction these expectation values determine the
representation up to a unitary equivalence (see e.g.\ \cite{br}),
the uniqueness follows. The unitary representation of the Lorentz
group with the desired properties (\ref{lor}) and (\ref{invac}) is
now obtained by
\begin{equation}
 U(\Lambda)W(D)R(c)\W_\ka
 =W(T_\Lambda D)R(T_\Lambda c)\W_\ka\,.
\end{equation}
One can easily show that the generators $M_{ab}$ of these
transformations, defined by $U(\delta^a{}_b+\w^a{}_b)\approx
\exp[-\frac{i}{2}\w^{ab}M_{ab}]$ for small antisymmetric
$\w_{ab}$, may be expressed as
\begin{equation}
 M_{ab}=-\frac{1}{4\pi}\int :\hat{c}(l)L_{ab}\hat{D}(l):\dl\,,
\end{equation}
where normal ordering is determined by point splitting as
\begin{equation}
 :\hat{c}(l)L_{ab}\hat{D}(l):=\lim_{l'\to l}
 \Big\{\hat{c}(l')L_{ab}\hat{D}(l)
 -(\W_\ka,\hat{c}(l')L_{ab}\hat{D}(l)\W_\ka)\id\Big\}\,,
\end{equation}
and the limit goes over $l'$ linearly independent from $l$.

The above construction leaves us with the freedom of one real
parameter~$\ka$ in the choice of representation. In the usual
situation for quantum fields the selection criterion which often
leaves only one representation is the demand that the vacuum state
be translation invariant and the total energy be a positive
operator. We do not have this criterion for our disposal in the
case of present model. However, Staruszkiewicz thinks that the
asymptotic field (\ref{asfE}) should ``remember'' that its first
term (the one explicitly written in (\ref{asfE})) is obtained from
the positive frequency field, which in usual electrodynamics
annihilates the vacuum. Thus he demands that the quantum version
of the first term in (\ref{asfE}) annihilates $\W_\ka$. Looking at
(\ref{asfE}), (\ref{repl}) and (\ref{covrep}) it is easy to
convince oneself that this condition is satisfied if, and only if,
\begin{equation}
 \ka={\frac{2}{\pi}}\,.
\end{equation}
In this way one arrives at an interesting and elegant model, which
explicitly depends on the value of elementary charge $e$ and has a
charged field among its variables. Staruszkiewicz believes, and in
fact this is his main motivation, that some mathematical and
physical consistency restrictions will squeeze out of this model
an information on the size of the fine structure constant
$e^2/\hslash c$. That this hope may, in fact, be justified, is
suggested by the structure of the Lorentz group representation
$U(\Lambda)$. As it turns out, the breakup of this representation
into irreducibles must depend nontrivially on the value of
$e^2/\hbar c$ \cite{sta3}.

We hope that the formulation of the Staruszkiewicz model we have
discussed here helps to clarify its structure at least for some
readers. But it should also help to simplify calculations. We give
as an example the calculation of the scalar product of states
$R(e/(v\cdot l)^2)\W_\ka$ (in Staruszkiewicz's notation
$e^{-iS_0}|0\rangle$ with $S_0$ the spherically symmetric part of
$S(x)$ in the reference system with time axis~$v$). Denote
$F_{v,u}(l)=e\log[v\cdot l/u\cdot l]$. Then by (\ref{weylst}) and
(\ref{vacexp}) we have
\begin{equation}
 \begin{split}
 &(R(e(v\cdot l)^{-2})\W_\ka,R(e(u\cdot l)^{-2})\W_\ka)
 =(\W_\ka,R(\D^2F_{v,u})\W_\ka)\\
 &=\exp\big[-(\ka/4)\|[F_{v,u}]\|_\K^2\,\big]
 =\exp{\big[-(e^2\ka/2)\big(\chi_{v,u}\coth\chi_{v,u}-1\big)\big]}\,,
 \end{split}
\end{equation}
where $v\cdot u=\cosh\chi_{v,u}$. For $\ka=2/\pi$ this reproduces
the result obtained in a much more involved way in \cite{sta1}. We
have used in the calculation:
\begin{equation}
 \begin{split}
 \|[F_{v,u}]\|_\K^2&=-\frac{1}{4\pi}
 \int[\D F_{v,u}(l)]^2\,\dl\\
 &=\frac{e^2}{4\pi}\int\Big[\frac{2\,v\cdot u}{(v\cdot l)(u\cdot l)}
 -\frac{1}{(v\cdot l)^2}-\frac{1}{(u\cdot l)^2}\Big]\,\dl\\
 &\hspace{0pt}=2e^2\bigg\{
 \frac{v\cdot u}{\sqrt{(v\cdot u)^2-1}}
 \log\Big[v\cdot u+\sqrt{(v\cdot u)^2-1}\Big]-1\bigg\}\,.
 \end{split}
\end{equation}

\section{Asymptotic causal algebra}\label{calg}

Let us now return again to the discussion of the asymptotic fields
considered in Section \ref{space}. Recall that the assumption of
their behavior as defined in (\ref{sppot}) was dictated by the
fall-off of Coulomb fields of charges. However, it later turned
out that one half of the resulting asymptotic fields, these of
magnetic type (\ref{asfM}), did not actually appear in real
processes, so they could be omitted. This left us with the
long-range characteristics of the electric type only. But now we
can ask further: do \emph{all} of these characteristics have a
role to play in real processes? Our answer is: no, and as we shall
see, this is precisely what allows us to construct an algebra
which unites both the usual local and the long-range degrees of
freedom.

The selection criterion for free electromagnetic fields we want to
use is this: we admit only those fields which may be produced as
radiation fields in processes involving scattering charged
particles or fields, asymptotically moving freely for early and
late times. Recall that radiation field is the difference between
the retarded and advanced field produced by the current. Take the
simplest instant of such field, the radiation field produced by a
charge scattered instantaneously at $x=0$. In this case the
radiation potential in spacelike directions is the difference of
two Coulomb fields
\begin{equation}
 A^\mathrm{rad}_b(x)=Q\bigg(\frac{v_b}{\sqrt{(v\cdot x)^2-x^2}}
 -\frac{u_b}{\sqrt{(u\cdot x)^2-x^2}}\bigg)\,,\quad
 x^2<0\,,
\end{equation}
where $Q$ is the charge of the particle, and $v$ and $u$ its
initial and final velocity respectively. Note that this potential
is homogeneous of degree $-1$, so its spacelike asymptotic limit
(\ref{sppot}) is given by the same function. More generally, if
the motion of the particle is modified but $v$ and $u$ remain its
asymptotic velocities, then the above formula still gives the
spacelike asymptotic $A^\as(x)$ of the potential. A striking
feature of this potential is its evenness:
\begin{equation}\label{even}
 A^\as(-x)=A^\as(x)\,,\qquad x^2<0\,.
\end{equation}
Now, this property is conserved under the superposition principle,
so it remains true for a general field produced by particles. One
can show that the same property holds for electromagnetic
potential radiated by scattered charged fields. Thus we take
(\ref{even}) as our selection criterion. Compare this with the
general asymptotic potential (\ref{aspot}). Our condition is then
equivalently expressed as
\begin{equation}
 \Ip a^\as_b(k)=0\,.
\end{equation}

We want to view our selection criterion from yet another
viewpoint. For a general Lorentz gauge potential of the form
(\ref{free}) let us denote for a future-pointing null vector $l$
and $s\in\mathbb{R}$:
\begin{equation}
 \V_b(s,l)=-\int \w a_b(\w l) e^{-i\w s}\,d\w\,,
\end{equation}
where dot denotes differentiation with respect to $s$. It is easy
to see that $\V_b(s,l)$ is a real function, orthogonal to $l^b$
and homogeneous of degree $-2$ in all its variables:
\begin{equation}
 l\cdot \V(s,l)=0\,,\qquad
 \V_a(\mu s,\mu l) = \mu^{-2} \V_a(s, l)\, ,~~\mu>0\, , \label{hom}
\end{equation}
A straightforward calculation then shows that the Fourier
representation (\ref{free}) may be written as
\begin{equation}\label{freedl}
 A_b(x)=-\frac{1}{2\pi}\int\V_b(x\cdot l,l)\,\dl\,.
\end{equation}
If $a_b(k)$ has a scaling limit (\ref{aas}) then taking into
account the reality condition (\ref{reality}) one finds that
 $\w\Rp a_b(\w l)$ is continuous in $\w=0$, while
 $\w\Ip a_b(\w l)$
has a jump of magnitude $2\Ip a^\as_b(l)$. This leads to the
estimate
\begin{equation}\label{vgen}
 \V_b(s,l)=-\frac{2}{s}\Ip a^\as_b(l)+O(|s|^{-1-\ep})\quad
 \text{for}\quad |s|\to\infty
\end{equation}
for some $\ep>0$. Now, consider the null asymptotics of the
potential, more precisely, take an arbitrary point in spacetime
$x$ and consider the asymptotics of $A(x+Rl)$ for $R\to\infty$.
One shows that if the leading term in (\ref{vgen}) does not
vanish, then the dominating term of this asymptotics is $2\Ip
a^\as_b(l)\log R/R$. As it turns out, in that case the leading
term for the angular momentum density at $x+Rl$ is of order $\log
R/R^2$. This means that even the \emph{differential} flux of
angular momentum radiated into infinity cannot be defined, which
is our second reason to reject those fields.

We want now to consider an interacting theory, and we take for
definiteness the classical theory of the electron-positron Dirac
field coupled by local gauge principle to the electromagnetic
field, with the intention of later ``quantization''. In
perturbative calculations one uses an approximation in which the
fields are free at very early and very late times, (matter is
completely decoupled from radiation). This procedure is assisted
by some preliminary regularization, such as restricting the
interaction to some subset of spacetime, which may be an effective
tool to do practical calculations, but is unable to satisfactorily
clarify the infrared structure. We want to improve on that
approximation so as to take into account the infrared degrees of
freedom and the Gauss law.

The selection criterion for the electromagnetic fields may still
be taken over to the interacting case in the following sense. If
$A_b$ is the Lorentz potential of the total field, then one
defines in standard way the incoming and outgoing free fields by
$A_b=A^\mathrm{ret}_b+A^\mathrm{in}_b=A^\adv_b+A^\out_b$, where
$A^\mathrm{ret}_b$ and $A^\adv_b$ are the retarded and the
advanced potential of the sources respectively. Then it may be
consistently assumed that both $A^\mathrm{in}_b$ and $A^\out_b$
satisfy the selection criterion.

Our aim is to consider fields at causally remote regions, ``in''
or ``out'', and we restrict attention to the ``out'' case. This is
usually taken to mean: on a spacelike hyperplane, which is taken
to the limit of time tending to $+\infty$. However, due to the
different propagation speeds of matter and radiation one can
exchange this for: matter field far away in the future timelike
directions, and electromagnetic field far away in the future null
directions. Consider the electromagnetic field first. With our
assumptions one shows that there is a function $V_b(s,l)$
homogeneous of degree $-1$ such that
\begin{equation}\label{nullas}
 \lim_{R\to\infty} RA_b(x+Rl)=V_b(x\cdot l,l)\,.
\end{equation}
This function is homogeneous of degree $-1$, satisfies
\begin{equation}
 l\cdot V(s,l)=Q\,,
\end{equation}
where $Q$ is the charge of the field, and is bounded by
\begin{equation}
 |\V_b(s,l)|\leq\frac{\con}{(t\cdot l)^2}
 \Big(1+\frac{|s|}{t\cdot l}\Big)^{-1-\ep}\,.
\end{equation}
(only the constant depends on $t$). The ``out'' field may be
recovered from this asymptotics by (\ref{freedl}), and its null
asymptotics is given by (\ref{nullas}) with $V_b(s,l)$ replaced by
$V^\out_b(s,l)=V_b(s,l)-V_b(+\infty,l)$. The limit value
$V_b(+\infty,l)$ is completely determined by the outgoing
currents, and determines according to (\ref{nullas}) the null
asymptotics of the advanced potential. The  spacelike asymptotics
of the ``out'' field is governed by
\begin{equation}
 a^\as_b(l)=-\frac{1}{2\pi}\int\V_b(s,l)\,ds
 =\frac{1}{2\pi}V^\out_b(-\infty,l)\,,
\end{equation}
but the spacelike asymptotics of the \emph{total} field is
determined by $V_b(-\infty,l)$, and for any point $x$ and
spacelike vector $y$ one has
\begin{equation}
 \lim_{R\to\infty}R^2F_{ab}(x+Ry)=\frac{1}{2\pi}
 \int\big(l_aV_b(-\infty,l)-l_bV_a(-\infty,l)\big)
 \delta'(y\cdot l)\,\dl\,.
\end{equation}
Note also, that the second and the third terms in the function
$S(x)$ as given by~(\ref{phase}) now vanish, so here one could not
construct an analogy of the Sta\-rusz\-kie\-wicz model -- function
$D(l)$ in (\ref{repl}) is identically zero. There is no need nor
space for the extension given by the first replacement in
(\ref{repl}) either. On the other hand, the constant in $\Rp a(l)$
\emph{will} appear in our model, and will be related to a phase
variable. We denote
\begin{equation}\label{phive}
 \Phi(l)=\frac{1}{4\pi}
 \int\frac{l\cdot V^\out(-\infty,l')}{l\cdot l'}\,\dl\,.
\end{equation}

Consider now the timelike asymptotics of the Dirac field
$\psi(x)$. One shows that with an appropriate choice of a local
gauge (locally related to the Lorentz gauge) one has for $v^2=1$,
$v$ future-pointing:
\begin{equation}
\p(\la v)\sim -i\la^{-3/2} e^{\txt -i(m\la+\pi/4)\g\cdot v}
f(v)~~{\rm for}~~\la\to\infty\, , \label{tas}
\end{equation}
where $\g^a$ are the Dirac matrices. Define, provisionally, the
free outgoing Dirac field by
\begin{equation}
\p^\out_f(x) = \Big(\frac{m}{2\pi}\Big)^{3/2} \int e^{\txt
-imx\cdot v\g\cdot v}\g\cdot v\,f(v) \m(v)\, ,
\end{equation}
where $d\mu(v)$ is the invariant measure $d^3v/v^0$ on the
hyperboloid $v^2=1$, \mbox{$v^0>0$}, and the formula is a concise
form of the Fourier representation of $\p^\out_f(x)$, reproducing
in the free field case the original field $\psi(x)$. The outgoing
current of the Dirac field is determined by $f(v)$, and one shows
that the lacking component $V_b(+\infty,l)$ of the total
electromagnetic potential is given by
\begin{equation}
 V_a(+\infty,l)=\int n(v) V^e_a(v,l)\, \m(v)\, ,
 \label{Gau}
\end{equation}
where $n(v)=\ov{f(v)}\g\cdot v f(v)$ is the asymptotic density of
particles moving with velocity $v$ and
\begin{equation}\label{coulomb}
 V_a^e(v,l) =ev_a/v\cdot l
\end{equation}
is the null asymptotics (\ref{nullas}) of the Lorentz potential of
the Coulomb field surrounding a particle with charge $e$ moving
with constant velocity $v$. Therefore, the above relation is the
implementation of the Gauss constraint on the space of classical
asymptotic variables.

The question now arises: do the fields $A^\out$ and $\p^\out_f$
separate completely in the ``out'' region? We interpret this
question as: can the total energy momentum and angular momentum of
the system be separated into contributions from $A^\out$ and
$\p^\out_f$? The answer is `yes' in the case of energy momentum,
but `no' in the case of angular momentum -- in this case there is
a term which couples the infrared degrees of freedom
$V^\out_b(-\infty,l)$ with $f(v)$. However, as it turns out, the
full separation may be achieved if one introduces a new variable
$g(v)$ by
\begin{equation}\label{dress}
 g(v)=\exp\Big(\frac{ie}{4\pi}\int
\frac{\Phi(l)}{(v\cdot l)^2}\,\dl\Big)\,f(v)\,,
\end{equation}
and defines the ``dressed'' free Dirac field by
\begin{equation}
\p^\out(x) = \Big(\frac{m}{2\pi}\Big)^{3/2} \int e^{\txt -imx\cdot
v\g\cdot v}\g\cdot v\, g(v) \m(v)\,.
\end{equation}
We draw attention of the reader to the following circumstances.
First, the transformation (\ref{dress}) is a very nonlocal one.
The asymptotics of the local Dirac field in the timelike direction
of $v$ is multiplied by a factor containing information on the
spacelike asymptotics of the outgoing electromagnetic field
$A^\out_b$. Next, as the conserved quantities have been completely
separated, the field $\p^\out$ should be regarded as describing
the charged particles \emph{together} with their Coulomb fields.
Finally, as announced earlier, the constant in $\Phi(l)$ does
appear in the model. However, this constant appears only in the
exponentiated form given by~(\ref{dress}). Thus we put forward the
interpretation
\begin{equation}\label{myphase}
 \frac{e}{4\pi}\int \frac{\Phi(l)}{(v\cdot l)^2}\,\dl
 =\text{phase variable}\,.
\end{equation}
Note that this definition involves only the free electromagnetic
characteristics, and is independent of particular matter field.

This classical asymptotic model has a natural ``quantization''
based on the heuristic demand that the total conserved quantities
generate Poincar\'e transformations. The model is formulated in
terms of the quantities which have direct physical meaning in the
asymptotic region, that is the asymptotics of the total field
$\hat{V}_b(s,l)$, and the asymptotics of the Dirac field with the
accompanying Coulomb fields of the particles $\hat{g}(v)$
(``hats'' indicate the quantum versions). We introduce the
following structures on the space of asymptotic variables: the
symplectic form
\begin{equation}
 \{ V_1, V_2\}
 =\frac{1}{4\pi}\int\big(\V_1\cdot V_2-\V_2\cdot V_1\big)(s,l)\,
 ds\,\dl\,,\label{sympl}
\end{equation}
and the scalar product
\begin{equation}
(g_1,g_2) = \int\ov{g_1(v)}\g\cdot v g_2(v) \m(v)\, , \label{sp}
\end{equation}
Let $g(v)$ and $V_b(s,l)$ be classical test fields describing
asymptotics of \emph{free} fields, thus, in particular,
$V_b(+\infty,l)=0$. The basic elements of the quantum model are
functionals of those test fields: $W(V)$ and $B(g)$. Loosely, one
can think of them as
\begin{equation}
 W(V)= e^{\txt -i\{V,\hat{V}\}}\,,
 \quad B(g)=(g,\hat{g})\,.
\end{equation}
Elements $W(V)$ and $W(V')$ are identified if the test potentials
$V_b(s,l)$ and $V'_b(s,l)$ give the same electromagnetic test
field asymptotics and the same phase variable (\ref{myphase}),
that is
\begin{equation}
 l_{[a}V'_{b]}(s,l)=l_{[a}V_{b]}(s,l)\,,\quad
 \Phi'(l)=\Phi(l)+n\frac{2\pi}{e}\,,
\end{equation}
where $\Phi(l)$ is related to $V_b(s,l)$ by (\ref{phive}). The
algebra is then defined by
\begin{equation}\label{asymalg}
 \begin{split}
 &W(V_1)W(V_2)=e^{\txt -\frac{i}{2}
 \{ V_1,V_2\}} W(V_1+V_2)\,,\\
 &\hspace{20pt}W(V)^*=W(-V)\,,\qquad W(0)=\id\,,\\
 &\hspace{-20pt}[B(g_1), B(g_2)]_+ =0\,,\quad
 [B(g_1), B(g_2)^*]_+ =(g_1, g_2)\id\,,\\
 &\hspace{20pt}W(V)B(g)=B(S_\Phi g)W(V)\,,
 \end{split}
\end{equation}
where
\begin{equation}
 \big(S_\Phi g\big)(v) = \exp\bigg(\dsp i\frac{e}{4\pi}
 \int\frac{\Phi(l)}{(v\cdot l)^2}\,\dl\bigg)\, g(v)\, .
\end{equation}
With a proper technical formulation of conditions on the scope of
test functions the above relations generate a $C^*$-algebra, which
I interpret as the algebra of asymptotic fields in quantum
electrodynamics.

The only relation in which the above algebra diverges from the
usual tensor product of independent algebras of the two fields
separately is the last relation in (\ref{asymalg}), but this is
the key to the physics of the model. We note that for the Coulomb
field asymptotics (\ref{coulomb}) one has
\begin{equation}
 \{V^e(v,.),V\}=
 \frac{e}{4\pi}\int\frac{\Phi(l)}{(v\cdot l)^2}\,\dl\,.
\end{equation}
The commutation relation between the fermionic operator $B(g)$ and
the electromagnetic operator $W(V)$ may be therefore written in
loose terms as
\begin{equation}
 e^{\txt-i\{V,\hat{V}\}}\hat{g}(v)
 =\hat{g}(v)e^{\txt-i\{V,\hat{V}-V^e(v,.)\}}\,.
\end{equation}
This means that the operator $\hat{g}(v)$, beside its fermionic
role which is to annihilate a particle with charge $e$ or create
one with the opposite charge, also annihilates or creates the
particle's Coulomb field respectively.

Within the model formulated here the following results are
obtained.
\begin{itemize}
 \item[(i)] The spectrum of the charge operator is quantized in units of
elementary charge. This is the consequence of the appearance of
the quantum phase. As this phase variable is tied to the free
electromagnetic potential, this quantization law is universal.
 \item[(ii)] In representations of the asymptotic algebra satisfying
Borchers' criterion (spacetime translations implementable by
unitary operators with the energy-momentum spectrum in the future
lightcone) the analogue of the functional form of Gauss'
constraint (\ref{Gau}) is satisfied.
 \item[(iii)] The importance of
the regularity of representations with respect to all Weyl
operators is stressed. The vacuum representation is shown to be
non-regular with respect to Coulomb field operators ($W(V)$ with
infrared singular test functions $V$), which leads to the loss of
the Coulomb field and to a nonphysical superselection structure. A
class of ``infravacuum'' representations is constructed, which are
``close to the vacuum'' but regular at the same time. Each
irreducible representation of the field algebra in this class
leads to the superselection structure of observables characterized
by the electric charge. There is neither a zero-energy vector
state nor mass-shell charged vector states in these
representations.
\end{itemize}

Finally, to make some contact with the Staruszkiewicz model again,
one can consider a kind of adiabatic limit (slowly varying fields)
of a Weyl model based on the symplectic form (\ref{sympl}) alone
(with no fermionic fields, but with charged test fields $V_b(s,l)$
admitted instead). That was done in \cite{her2}. The mathematics
of the resulting model is identical with that of Staruszkiewicz's
model, and in fact our formulation of the latter as a kind of Weyl
algebra given in Section \ref{stam} was based on that paper.
However, the interpretation of variables is different in the two
cases. In particular, the quantity (\ref{myphase}) survives the
adiabatic limit as a phase variable, which is different from
Staruszkiewicz's phase.

\end{document}